\documentclass[fleqn,usenatbib]{mnras}

\usepackage[T1]{fontenc}

\DeclareRobustCommand{\VAN}[3]{#2}
\let\VANthebibliography\thebibliography
\def\thebibliography{\DeclareRobustCommand{\VAN}[3]{##3}\VANthebibliography}

\usepackage{graphicx}	
\usepackage{amsmath}	
\usepackage{amssymb}	
\usepackage{newtxtext,newtxmath}
\usepackage{xcolor}
\newcommand{\modfm}[1]{\textcolor{red}    {#1}}

\definecolor{royalazure}{rgb}{0.0, 0.22, 0.66}

\usepackage[normalem]{ulem}

\title[Sandwiched planet formation]{Sandwiched planet formation: restricting the mass of a middle planet}

\author[Matthew Pritchard et al.]{
Matthew Pritchard,$^{1,2}$
Farzana Meru,$^{1,2}$\thanks{E-mail: f.meru@warwick.ac.uk},
Sahl Rowther$^{1,2,3}$,
David Armstrong$^{1,2}$
and Kaleb Randall$^{1,2}$
\\
$^{1}$Department of Physics, University of Warwick, Gibbet Hill Road, Coventry, CV4 7AL, UK\\
$^{2}$Centre for Exoplanets and Habitability, University of Warwick, Gibbet Hill Road, Coventry CV4 7AL, UK\\
$^{3}$School of Physics and Astronomy, University of Leicester, Leicester LE1 7RH, UK
}

\date{Accepted XXX. Received YYY; in original form ZZZ}

\pubyear{2021}

\begin{document}
\label{firstpage}
\pagerange{\pageref{firstpage}--\pageref{lastpage}}
\maketitle

\begin{abstract}

We conduct gas and dust hydrodynamical simulations of protoplanetary discs with one and two embedded planets to determine the impact that a second planet located further out in the disc has on the potential for subsequent planet formation in the region locally exterior to the inner planet. We show how the presence of a second planet has a strong influence on the collection of solid material near the inner planet, particularly when the outer planet is massive enough to generate a maximum in the disc's pressure profile. This effect in general acts to reduce the amount of material that can collect in a pressure bump generated by the inner planet. When viewing the inner pressure bump as a location for potential subsequent planet formation of a third planet, we therefore expect that the mass of such a planet will be smaller than it would be in the case without the outer planet, resulting in a small planet being sandwiched between its neighbours -- this is in contrast to the expected trend of increasing planet mass with radial distance from the host star. We show that several planetary systems have been observed that do not show this trend but instead have a smaller planet sandwiched in between two more massive planets. We present the idea that such an architecture could be the result of the subsequent formation of a middle planet after its two neighbours formed at some earlier stage.
\end{abstract}




\section{Introduction}



The discovery of the Kepler planets has hugely increased the number of observed exoplanets, and with that our ability to statistically understand the properties of planets and the architectures of planetary systems has grown.  One property that is particularly intriguing from a planet formation perspective is the mass ordering of neighbouring planets.  Given that the majority of planets known to date have been discovered through the transit technique, most of the data on neighbouring planets in a system appears to be on their radii.  \citet{Weiss2018} analysed a set of 909 Kepler planet candidates from the California Kepler Survey from 355 multi-planet systems and found that in general, neighbouring planets seem to be reasonably similar in size with a trend towards larger planets existing at larger orbital separations in $65\%$ of the cases (also see \citealp{Lissauer_Kepler}, \citealp{Ciardi_Rp_trend} and \citealp{Chevance_massordering2021}). 
 The former trend has been termed \emph{Peas in a Pod} (see \citealp{Weiss_PiaP_PPVIIreview} for a recent review) and it has been suggested that this phenomenon likely occurs from birth \citep{Chevance_massordering2021}.  However the Peas in a Pod trend is still under debate given that it is thought that giant planets further out than what can be observed by Kepler are thought to exist with smaller inner planets \citep{Zhu_Wu_coldJup_superEarth,Bryan_coldJup_superEarth,Herman_coldJup_superEarth} as well as the claim that the Peas in a Pod trend may suffer from detection biases \citep[][]{Zhu2020,Murchikova2020} (though this has also been contested by \citealp{Weiss_Petigura2020}).  This trend may also be affected if small, undetected planets exist in these systems \citep[][]{Zhu2020}.

Understanding the mass trends are more challenging.  \citet{Millholland2017}  analyse 55 multiplanet systems consisting of a total of 145 Kepler planets whose masses were determined using TTVs \citep[][]{Hadden2017} and showed a similar Peas in the Pod trend for planet mass, whereby neighbouring planets are similar in mass (compared to a randomised distribution of planets) and that in a single system they are ordered in mass so that the outer planets are more massive (also see \citealp{Bashi_Zucker2021} and \citealp{Chevance_massordering2021}, as well as \citealp{Goyal_Wang_nonTTVs_PiaP} for non-TTV planets).  However, \citet{Zhu2020} also argues that these results may be due to observational biases.  More recently, a theoretical study by \citet{Mishra2021} considered whether the Peas in the Pod trends exist in the results of population synthesis models of planet formation and evolution, that are intended to be mock Kepler observations.  They find that neighbouring planets do have similar masses and that 55\% of neighbouring planets are ordered in mass.  Nevertheless this means that a significant proportion of neighbouring planets do not adhere to this mass ordering trend (also see \citealp{Mishra2023}).

From a planet formation perspective, an increase in planet mass with orbital distance is indeed expected: the isolation mass -- which is the amount of solid material that can be fed to the planet before the dust supply is reduced due to the planets' changing gas surface density (and hence pressure profile) -- in a protoplanetary disc increases with radial distance from the star \citep[][]{Lambrechts2014, Bitsch2018, Ataiee2018}.  In addition, the Hill radius also increases with increasing orbital separation, which means that even for similar mass planets, the amount of material available in an outer planet's feeding zone is larger due to an increase in the local disc mass, $M_{\rm d,local} \sim \Sigma R^2$, at larger radii.  Moreover, the inside-out planet formation model also leads to an increase in planet mass with distance \citep[][]{Chatterjee2014}.  In this model dust collects into a ring and grows into a planet at the pressure maximum at the inner edge of the dead zone, after which the planet opens up a gap causing the dead zone to retreat and for the cycle to continue at larger disc radii.

Despite this, there are some ways in which planetary architectures may not follow this typical trend.  From an observational perspective, the NASA exoplanet archive\footnote{https://exoplanetarchive.ipac.caltech.edu/} shows a substantial fraction of cases where a planet in between two other planets does not follow the trend of increasing mass with increasing orbital distance.  We discuss these in more detail in Section~\ref{sec:observations}.

Generally sequential planet formation is thought to occur from inside-out due to the timescales involved in the core accretion process, or naturally through the inside-out planet formation process \citep[][]{Chatterjee2014} described above.  However in this paper we explore what happens to the dust flow inwards in a protoplanetary disc when two planets have already formed, and the impact that the outer planet can have on the collection of dust in the ring just exterior to the inner planet.  We present the idea of a sandwiched planet formation process -- whereby two planets form initially in a disc and then we hypothesise if a third (smaller) planet has some chance of forming in between the two, with a mass that does not follow the expected trend of increasing mass with increasing orbital distance.

How might such a process present itself in the formation process whereby two planets have already formed before the intermediate third one?  The inner planet may form as expected from the aforementioned traditional view of sequential planet formation.  We may then hypothesise that a second planet may form either preferentially at a special location such as an ice line \citep{Ros_Johansen_iceline,Cridland_dust_iceline_deadzone}, or that a second planet may form exterior to the first and migrate outwards \citep[e.g. to a heat transition or ice line;][]{Hasegawa_Pudritz_planettraps,Speedie_bifurcation}.  In these cases the dust flow into and out of the region in between the two planets will be determined by the planetary masses.  Alternatively if planets were to form quickly by Gravitational Instability (which may produce Neptune-mass planets if the discs are magnetised; \citealp{Deng_mag_GI_Neptune}), we may find ourselves with a similar scenario of two planets having formed with space in between for dust evolution to occur.  Finally, whilst not thoroughly explored, the idea of Core Accretion occurring in a gravitationally unstable disc may also lead to the formation of multiple planets, which could potentially be low in mass depending on the efficiency of this process.

In this work we make no assumptions about the way in which the two initial planets form but simply assume that they have formed, and observe the resulting dust flow that occurs due to the second planet being present.  In Section \ref{sec:methods} we present our method.  In Sections \ref{sec:results} and \ref{sec:discussion} we present the results of our simulations and discuss them, respectively.  Finally we present our conclusions in Section \ref{sec:conclusions}.

\section{Methods}
\label{sec:methods}

We perform two-dimensional  hydrodynamical simulations of gas and dust in protoplanetary discs with embedded planets, using the {\sc fargo3d} code \citep[][]{FARGO3D}.  Due to the isothermal equation of state, this is a scale-free code whose variables need to be rescaled to give  physical quantities. We treat dust species as pressureless fluids, an approximation that is valid for Stokes numbers $\rm St\lesssim1$; dust coagulation and fragmentation as well as back-reaction onto the gas are neglected, thus all dust species are independent. Dust diffusion is included with a Schmidt number $\rm Sc = 1$. This is largely the same code as that used by \citet{Rosotti2016} and \citet{Meru2019}, with the exception of slightly different boundary conditions that we describe below.

The disc is simulated on a $N_{\phi}\times N_{R}$ cylindrical grid with $N_{\phi}=1024$ and $N_{R}=638$, with logarithmically spaced grid cells in the radial direction to ensure square cells across the computational domain. This resolution resolves the Hill radius of the inner and outer planets by 7-8 and 8-10 grid cells, respectively, in both the radial and azimuthal directions. In code units, the inner and outer disc boundaries are located at $R=0.2$ and $R=10$, respectively, and we ensure that the choice of the inner boundary does not cause significant artefacts in the inner disc (see Appendix~\ref{sec:Numerical tests} for a comparison with an inner boundary of $R=0.5$).  Closed (anti-symmetric) boundary conditions are applied to the gas radial velocity, with gas not being allowed to flow through the inner or outer boundaries. The gas density is extrapolated according to a Keplerian power law at both the inner and outer boundaries.  We apply Stockholm wave-killing zones to the gas at both boundaries \citep[][]{deValBorro2006}, but with the locations of the wave-killing boundaries as described by \citet{McNally2019}.  We also ensure that the 2:1 resonant locations for each planet are contained within the undamped region of the disc \citep[][]{Benitez-Llambay2016}. Note that the damping is only applied to the radial velocity of the gas in order to conserve mass and angular momentum as in \citet{Dempsey2020}.
Inner and outer boundaries are open for the dust (using a symmetric boundary condition for the dust radial velocity), simulating dust flow from the outer disc through the outer boundary and from the simulated region to the inner disc through the inner boundary. The dust density is held fixed at the outer boundary according to the initial dust distribution and extrapolated at the inner boundary according to a Keplerian power law.

In all the simulations, the planets do not feel the gravitational potential of the other planet; and for all simulated systems, there is no overlap of planetary Hill spheres and hence planet-planet interactions are expected to have a limited effect. The planets are introduced into the disc simultaneously and do not accrete material. Furthermore, planet migration is not included, and the planets are held on fixed circular orbits.

\subsection{Disc Parameters}
\label{sec:disk params}

We use the same disc surface mass density and temperature parameters as that used by \citet{Rosotti2016} and \citet{Meru2019}.  The disc's gas distribution is modelled as:
\begin{equation}
    \Sigma=\Sigma_{0}\left(\frac{R}{R_{0}}\right)^{-1},
    \label{eq:density}
\end{equation}
where $\Sigma$ is the vertically-integrated gas density with $\Sigma_0 = 1 \times 10^{-3}$ and $R_0 = 1$ in code units is a normalisation constant.  These simulations are scale-free but as an example, for a $1 M_{\odot}$ star and $R_0 = 10$~au, $\Sigma_0$ would be $89\rm g cm^{-2}$ and the disc mass simulated would be $0.06M_{\odot}$.  The disc's aspect ratio varies according to
\begin{equation}
    \left(\frac{H}{R}\right)=0.05\left(\frac{R}{R_{0}}\right)^{0.25}.
    \label{eq:aspect}
\end{equation}

We choose a fiducial viscosity parameter \citep[][]{ShakuraandSunyaev1973} of $\alpha=10^{-3}$ for our main body of simulations. To evaluate the influence of viscosity on our conclusions, additional simulations are performed with $\alpha=10^{-4}$ (see Appendix \ref{sec:viscosity}).

We simulate three dust species with logarithmically-spaced Stokes numbers between 0.2 to $2\times10^{-3}$. The initial dust distribution has the same power law profile as that of the gas.  Given that we simply model how the dust moves in response to the gas without any back-reaction, the normalisation of the dust density is arbitrary; in what follows we assume a canonical value of the dust-to-gas ratio of 0.01.

\subsection{Suite of simulations}

\begin{figure*}
    \centering
    \includegraphics[width=\textwidth]{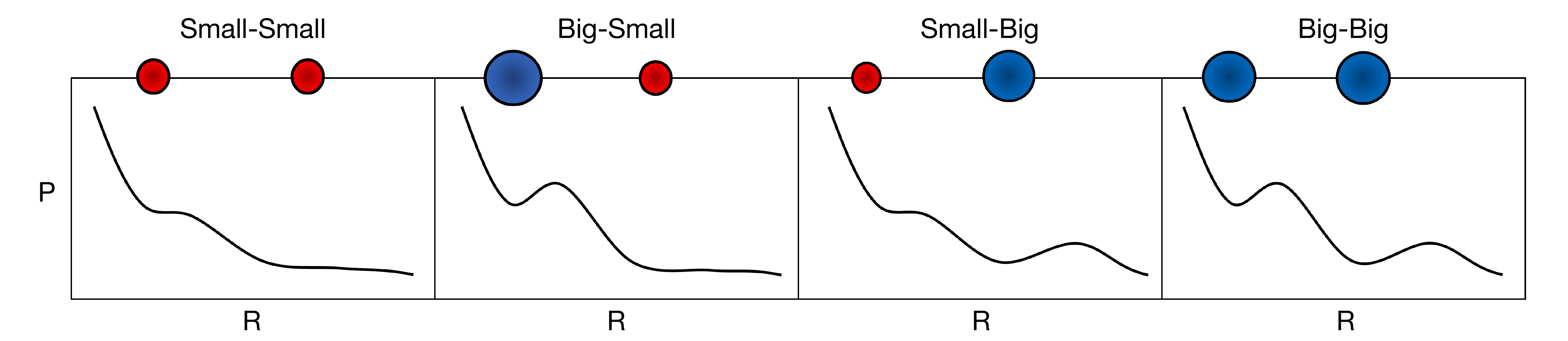}
    \caption{Schematic illustration of disc pressure profiles in the four cases we explore in our main suite of simulations.  "Small" and "Big" refer to low  and high mass planets, which cause the pressure profile exterior to the planet to form a point of inflection and a pressure maximum, respectively.}
    \label{fig:schematic}
\end{figure*}

\begin{table*}
	\centering
	\caption{Table of Simulations. The names Small-Small etc. correspond to the four cases of influence on the disc pressure profile illustrated in Figure~\ref{fig:schematic}. Whether or not a given planet generates a gas pressure maximum is indicated; this can be seen in the last row of Figure \ref{fig:main_fig}, where a pressure maximum corresponds to a region where $d {\rm ln} P/ d {\rm ln} R > 0$.}
	\label{tab:simulations_table}
	\begin{tabular}{lccccr} 
		\hline
		Simulation & \multicolumn{2}{c}{Inner planet}&\multicolumn{2}{c}{Outer planet}&$\alpha$\\
		& Mass ($M_{\oplus}$) & Pressure maximum? & Mass ($M_{\oplus}$) & Pressure maximum? &\\
		\hline
		\hline
		Small-Small & 12 & No & 20 & No & $10^{-3}$\\
		Big-Small & 20 & Yes & 20 & No & $10^{-3}$\\
		Small-Big & 12 & No & 35 & Yes & $10^{-3}$\\
		Big-Big & 20 & Yes & 35 & Yes & $10^{-3}$\\
		\hline
	\end{tabular}
\end{table*}

We focus on four cases corresponding to inner-outer planet mass combinations where each planet has a weak or relatively strong influence on the the disc's gas pressure profile; i.e. either generates a point of inflection or a pressure maximum. These four cases are illustrated schematically in Figure \ref{fig:schematic}. The planets are placed at $R=1$ and $R=2$ (code units). Assuming a stellar mass of $1M_{\odot}$, the chosen inner planet masses correspond to 12$M_{\oplus}$ and 20$M_{\oplus}$ (equivalent to planet-to-star mass ratios of $q = 3.6 \times 10^{-5}$ and $6 \times 10^{-5}$, respectively) resulting in a point of inflection and a maximum in the disc's pressure profile, respectively. For the outer planet, we simulate masses of 20$M_{\oplus}$ and 35$M_{\oplus}$ (equivalent to $q = 6 \times 10^{-5}$ and $1 \times 10^{-4}$, respectively). Each of our main simulations were run for 3000 orbits and the low viscosity simulations in Appendix~\ref{sec:viscosity} were run for 9000 orbits (at the inner planet's location\modfm). For comparison purposes, we also conduct single-planet simulations for each system, i.e. simulations of just the inner planet alone. We note that the single 12$M_{\oplus}$ and 20$M_{\oplus}$ planet cases reproduce the simulations of \citet{Rosotti2016} and give similar results. The two-planet simulations are summarised in Table~\ref{tab:simulations_table}.

In addition to the further simulations with $\alpha=10^{-4}$ mentioned above in Section \ref{sec:disk params}, we also conduct additional test simulations with varied resolution and outer boundary locations, and find that this has no effect on the conclusions presented. Changing the inner boundary position did have an effect on the disc profiles, however this was only in the inner disc. For more details see Appendix \ref{sec:Numerical tests}.

\section{Results}
\label{sec:results}

\begin{figure*}
	\includegraphics[width=\textwidth]{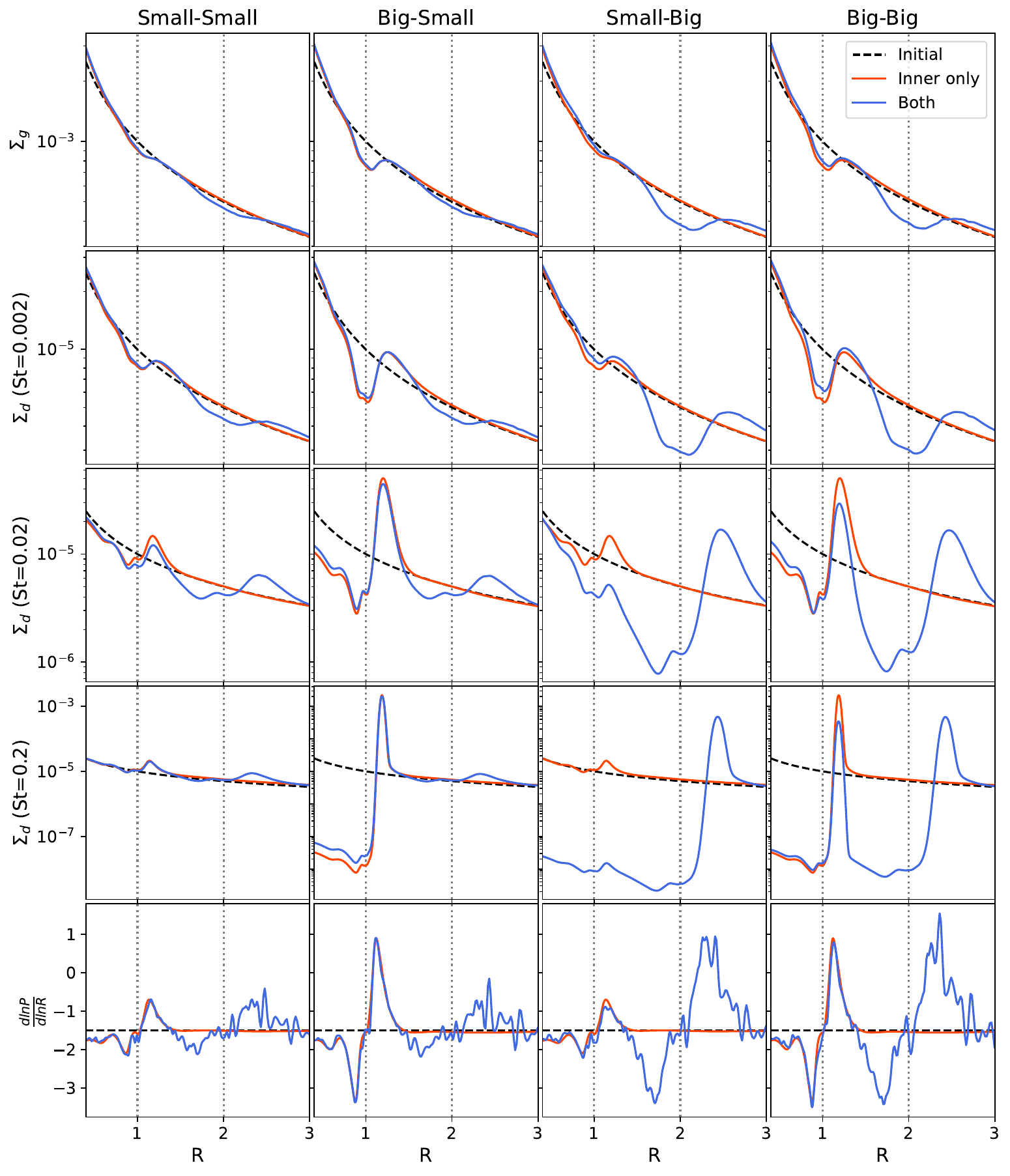}
    \caption{Azimuthally averaged profiles of the four simulations presented in Table~\ref{tab:simulations_table} with two planets that cause two points of inflections (Small-Small; left), an inner pressure maximum and an outer point of inflection (Big-Small; second column), an inner point of inflection and an outer pressure maximum (Small-Big; third column) and two pressure maxima (Big-Big; right).  The location of the two planets are represented by the vertical dotted lines.  The simulations with only a single planet (located at $R = 1$) are shown in a red line for comparison.  From top to bottom we present the gas surface density, the dust surface density for $\rm St=0.002$, $\rm St=0.02$ and $\rm St=0.2$ and finally the pressure gradient through the quantity $d {\rm ln} P/ d {\rm ln} R$, all in code units.  In cases with a small outer planet (columns 1 and 2) little difference is seen in the interplanetary density profiles (rows 1-4) between the one- and two-planet cases, whereas the difference is substantial for a large outer planet.  In particular, for the case where there are two planets present that are massive enough to create two pressure maxima (right column) the amount of dust in the pressure maximum in between the two planets is reduced.}
    \label{fig:main_fig}
\end{figure*}

\begin{figure}
    \centering
    \includegraphics[width=\columnwidth]{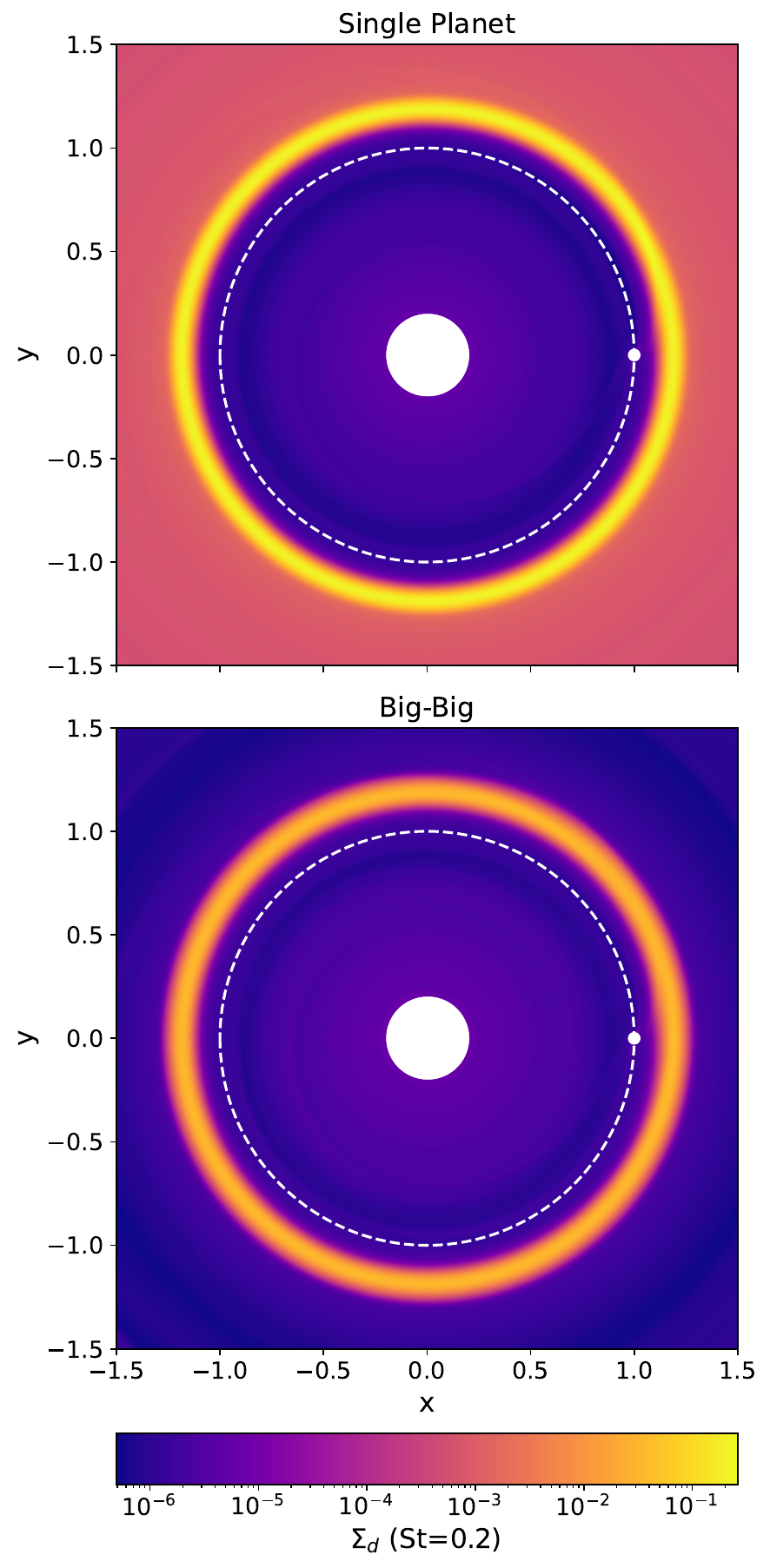}
    \caption{2D plot of the $\rm St = 0.2$ dust density for the single $20M_{\oplus}$ and the two-planet $20M_{\oplus} \& 35M_{\oplus}$ (i.e. Big-Big) simulations showing only the inner part of the discs simulated. The inner planet's position is indicated with a white circle, its orbit by the white dotted line. The high density ring exterior to the inner planet's orbit is a potential location for planet formation. In the two-planet case the density in this ring is notably reduced, limiting the mass of any compact bodies that may form here.}
    \label{fig:2Dfig}
\end{figure}

\begin{figure}
    \centering
    \includegraphics[width=\columnwidth]{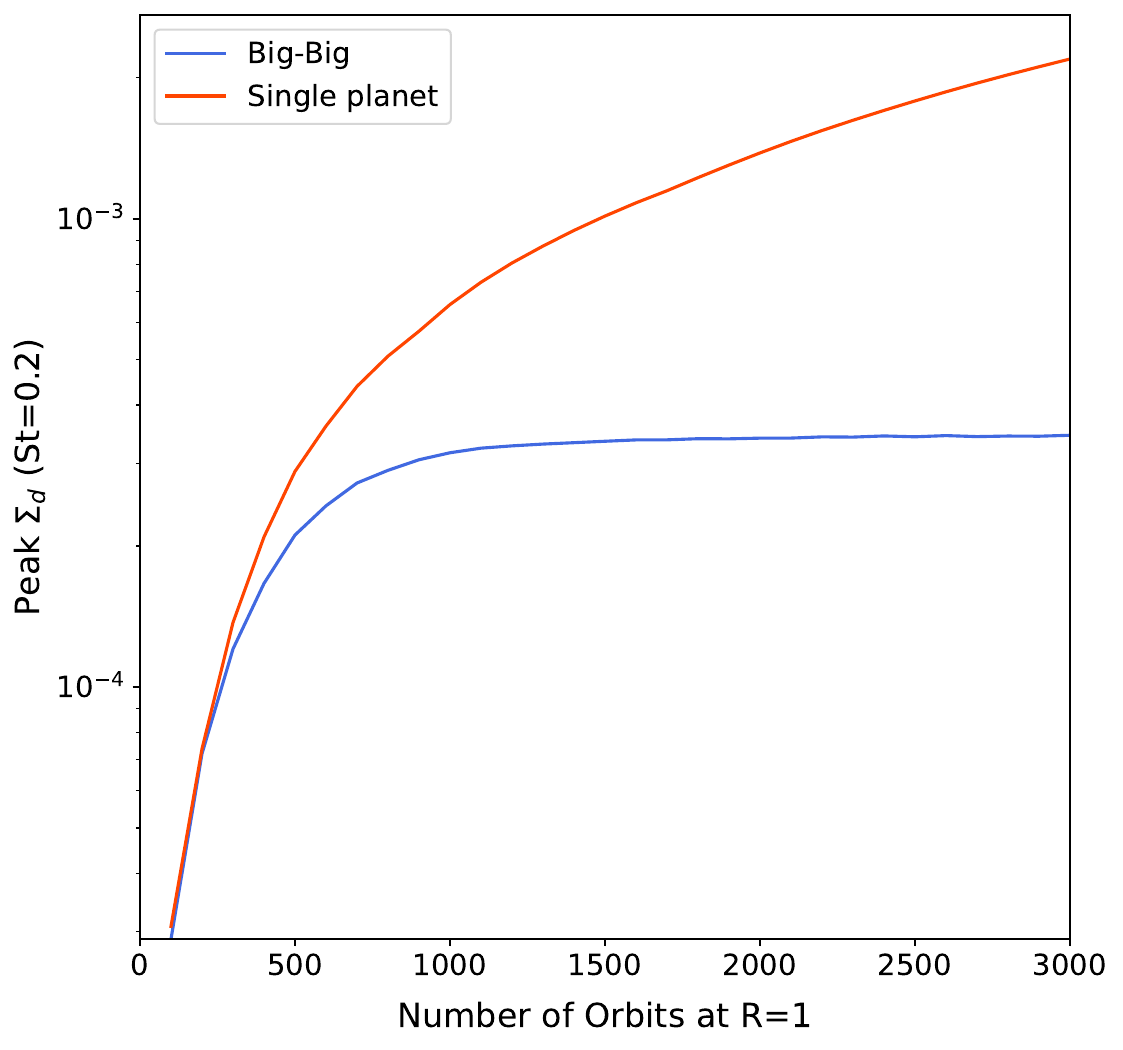}
    \caption{The maximum dust density (in code units) of large grains (St = 0.2) in the pressure maximum created by the planet at $R=1$. The peak density continues to increase for the single planet case. Whereas with an additional big outer planet, the peak density is restricted.}
    \label{fig:timeEvol}
\end{figure}

Fig.~\ref{fig:main_fig} summarises the results of our simulations by comparing the disc profiles between the single- and two-planet cases. Each row corresponds to the azimuthally averaged profiles for the gas surface mass density, the dust surface mass densities for each dust species and the quantity $d {\rm ln} P/ d {\rm ln} R$, for each of the four cases described above. The latter quantity describes the pressure gradient such that the presence of positive values indicates that a pressure maximum exists, while changes in negative values results in a point of inflection.  For these low-mass planets, the effect on the disc profiles are restricted to the regions locally interior or exterior to the planet, with the profiles being close to the initial unperturbed profiles in the outer parts of the disc.

We present the results associated with each of the scenarios referred to in Figure~\ref{fig:schematic}, though Section~\ref{sec: big-big} presents the main results associated with the Sandwiched Planet Formation that we present in this paper.

\subsection{Small-Small (two points of inflection)}
\label{sec: small-small}
In the single planet case with only the inner planet, the traffic jam caused by the small influence on the disc pressure profile results in only a modest build up of dust, with the local dust density increasing by a factor of a few compared to the initial value for the largest dust. The outer planet's small influence on the gas pressure profile also only results in a modest dust enhancement exterior to it. Crucially, the outer planet's small pressure perturbation has a very small impact on the flow of dust into the inner disc. As a result, the two-planet disc profiles (blue lines in the left column of Fig.~\ref{fig:main_fig}) in the interplanetary region and close to the inner planet show only small differences from the single inner planet case (red lines).

We note that the reduction in the peak for $\rm St = 0.02$ is greater than for $\rm St = 0.2$.  At the time plotted, the amount of dust in the peak has not reached a steady-state and continues to increase slowly over time.  However, the conclusion that the low mass outer planet only has a small impact on the dust flow remains unchanged.  As seen in Section~\ref{sec: big-big}, the main focus of this paper is the impact that two massive planets have on the architecture of the planetary system.  Therefore we do not explore this reduction in detail, but note that a future study that quantifies the impact of a small outer planet would be needed to consider this in more detail.

\subsection{Big-Small (pressure maximum and point of inflection)}
\label{sec: big-small}
In this case, the inner planet now has a significant impact on the disc profile. Generating a pressure maximum, the inner planet causes the formation of a dust trap exterior to its orbit. With continued inflow from the outer disc, significant dust collection results, with peak increases of two orders of magnitude for the largest dust (see the second plot in the fourth row of Fig.~\ref{fig:main_fig}). As in Section \ref{sec: small-small}, the outer planet's small influence on the gas pressure profile has little effect on the amount of dust that can drift past it into the interplanetary region, hence the flow of dust into the dust trap is largely unchanged, resulting in a similar dust disc profile in the region exterior to the inner planet to that of the single inner planet case. This is seen by comparing the red lines (inner planet only) to the blue lines (two-planet case) in the second column of Fig.~\ref{fig:main_fig}.

\subsection{Small-Big (point of inflection and pressure maximum)}
\label{sec: small-big}
In this case it is the outer planet that generates a pressure maximum and the inner planet that creates a traffic jam effect. The resulting dust trap in the outer disc greatly reduces the amount of dust that can drift past the planet and into the interplanetary region, resulting in a large dust depletion compared to the initial profile. For the largest two Stokes numbers, this effect greatly overshadows the modest dust enhancements the inner planet causes in the single-planet case. To see this, compare the red line (inner planet only) with the blue line (both planets) in the third and fourth plots in the third column of Fig.~\ref{fig:main_fig}. The small dust enhancement exterior to the inner planet is still seen in the two-planet case, but this degree of enhancement is much smaller than the overall level of depletion in the interplanetary region and inner disc. Interestingly, a small enhancement in the surface density of the smallest species of dust ($\rm St = 0.002$) is seen in the inner pressure bump (see second plot in the third column of Fig.~\ref{fig:main_fig}), which may well be due to the outer planet pushing more gas into the inter-planetary region.

\subsection{Big-Big (two pressure maxima): Sandwiched planet formation}
\label{sec: big-big}

As in Section \ref{sec: small-big}, the outer planet causes a significant dust depletion in the region interior to its orbit due to it’s pressure maximum stopping the dust in the outer disc from drifting inwards. This also causes the amount of dust collected in the trap created by the inner planet to decrease compared to the single planet case for the largest two dust sizes (and especially for the largest dust size; also see Figure~\ref{fig:2Dfig}).  We again see a minor enhancement in the density profile of St=0.002 dust in the inner dust trap compared to the inner-only case which again, may be attributed to the outer planet pushing more gas (and hence the small sized dust since it is more coupled to the gas) into the inter-planetary region.

We also note that the restricted amount of dust in between the two planets is evident at all times.  Figure \ref{fig:timeEvol} shows that the amount of large dust in the pressure maximum increases over time for the single planet case, whereas with an additional planet the maximum density in the ring levels off.

\section{Discussion}
\label{sec:discussion}

\subsection{Potential consequences for sequential planet formation}
\label{sec:implications}

Section~\ref{sec:results} shows that when the outer planet is small enough such that the perturbed pressure profile only creates a traffic jam, the impact on the inner dust collection is largely unaffected.  As expected, the impact on the inner dust collection happens only when the pressure profile in the outer disc due to the second planet is such that a pressure maximum forms.  In this case, if the inner planet is low in mass such that it cannot form a pressure maximum, the inner disc is depleted of its larger sized particles, consistent with past findings that explore dust filtration \citep{Zhu_dust_filtration,Rice_dust_filtration}.

However the more interesting aspect related to planet formation is when both planets are massive enough to create a pressure maximum.  In this case dust cannot be lost into the inner disc, but instead a depletion in the amount of dust in the inner ring occurs.  Depending on the amount of dust that is collected in this ring, there could potentially be enough material to form a low mass planet in between two higher mass planets. This therefore gives a pathway for the formation of planetary systems with planets in a Big-Small-Big mass pattern -- we refer to this as 'sandwiched planet formation'.

This is in contrast to the expectation that higher mass planets should form in the outer disc due to a larger isolation mass, a larger Hill radius, or the inside-out planet formation model \citep[][]{Chatterjee2014}, but does describe a subset of observed planets whose masses have been determined that do not follow the typical trend of increasing planet mass with radial location.

\subsection{Comparison with exoplanet observations}
\label{sec:observations}

\begin{figure}
    \centering
    \includegraphics[width=\linewidth]{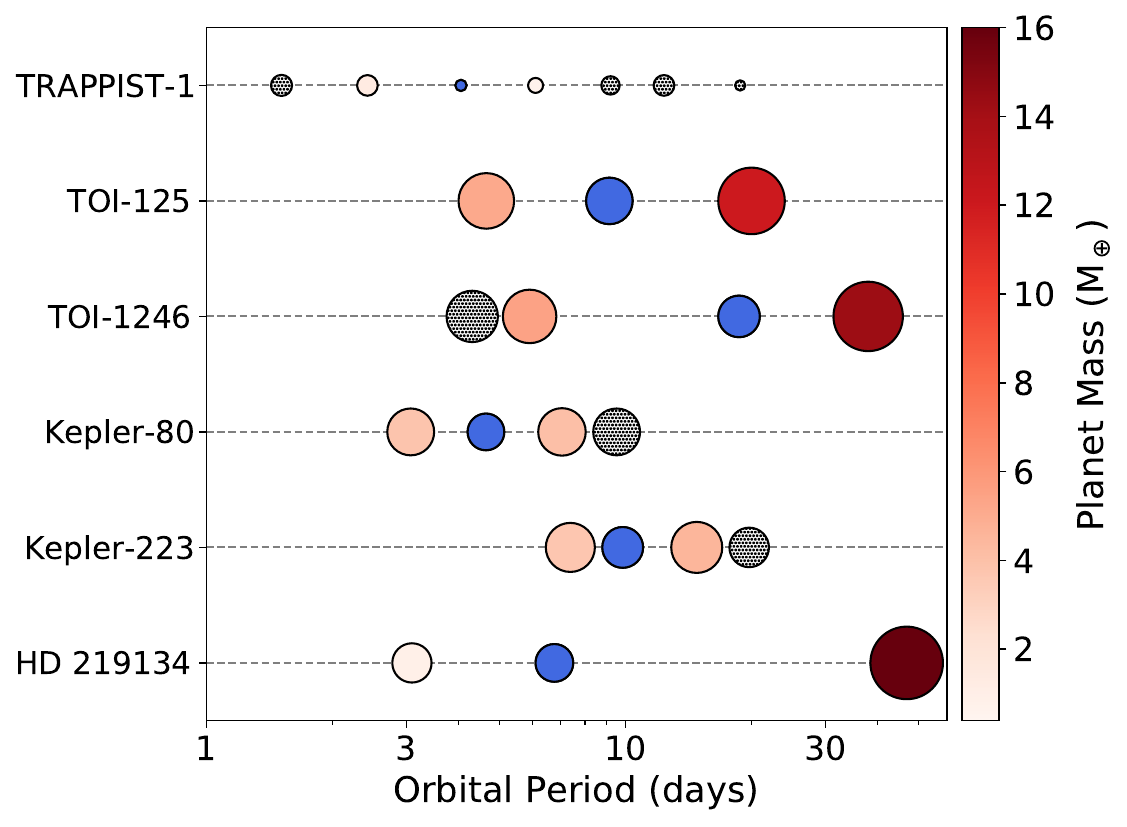}
    \caption{The 6 planetary systems which contain a small planet sandwiched in between two larger planets. The sandwiched planet is highlighted with a blue marker. The size represents the planet's mass as does the colour of the planet either side of the sandwiched planet.  The black hatched markers are the other planets in the system.}
    \label{fig:SW_Planets}
\end{figure}

To compare with observations, we only consider planetary systems from the NASA Exoplanet Archive\footnote{https://exoplanetarchive.ipac.caltech.edu} that satisfy certain criteria. Any planets that have been flagged as `controversial' have not been included. Planets without measurements of both the orbital period and mass are excluded.  To avoid complications arising from degeneracies between planet mass and inclination, we start by only considering systems with at least 3 transiting planets with known radii which we call the 'transiting requirement', and with masses determined to at least $3\sigma$ precision. As of April 25$^{\text{th}}$ 2023, there are only 17 such planetary systems known. A planet is considered to be small if the mass difference relative to both its neighbours satisfies the following condition,
\begin{equation}
\label{eq:massDiff}
    \frac{M_{i\pm1} - M_\mathrm{i}}{\delta\left( M_{i\pm1} - M_\mathrm{i} \right)_\mathrm{lower}}  > 1,
\end{equation}
where $M$ is the planet's mass. The subscripts $i$, and $i\pm1$ refer to the middle planet, and the planets exterior and interior to the middle planet, respectively. The denominator is the lower uncertainty on the mass difference of the middle planet and its neighbour. Equation \ref{eq:massDiff} is calculated using the method described in Appendix B in \cite{2019Laursen}, which takes into account the asymmetric uncertainties of the planets' mass measurements.  Note that the denominator involves using the lower uncertainty on the mass difference as this is the one that determines whether the middle planet is larger or smaller than the outer planets.  Of the 17 planetary systems, 6 of these have a clearly lower mass planet in the middle of the chain surrounded by higher mass planets as shown in Figure \ref{fig:SW_Planets}. These systems are HD 219134 \citep{2015VogtHD219134,2017GillonHD219134}, Kepler-80 \citep{Macdonald_Kepler80,2018ShallueKepler80}, Kepler-223 \citep{Mills_Kepler223}, TOI-125 \citep{Nielsen_TOI125}, TOI-1246 \citep{2022TurtelboomTOI1246}, and TRAPPIST-1 \citep{2021AgolTrappist1}.  The blue markers represent the smaller planet sandwiched by two bigger planets. The size of the markers represents the planet's mass, as does the colour of the two bigger planets. The black hatched markers show the other planets in the system.  If the 'transiting requirement' is ignored, 19 out of 72 planetary systems contain a lower mass planet sandwiched between two more massive planets.

We now consider the systems with only two detected transiting planets.  As observations are biased towards detecting the inner-most planets, it is possible that there is an additional undetected outer planet could potentially result in a sandwiched planetary system.  Using the same criteria as above, there are 34 two-planet systems. Of these, 6 systems are in the Big-Small configuration, i.e. where the inner planet is more massive.  If this unknown outer planet is larger than the small planet, then these 6 systems could follow the Big-Small-Big mass ordering. If the 'transiting requirement' is relaxed, there are 33 systems out of 193 that could contain a sandwiched smaller planet. Similarly, if we now consider 3+ planet systems where the two outermost detected planets are in the Big-Small configuration, 15 out of 43 such systems could contain a sandwiched smaller planet. Relaxing the 'transiting requirement' makes it 34 potential systems out of 98.

The population of known planets is subject to extensive biases, as a result of instrument capabilities, observer choices and stellar properties. Additionally, multiple detectable planets in-system produces complications, encouraging further observations due to the increased scientific value while making the modelling of such systems more challenging and increasing the amount of data required.  Furthermore, the detection method will create biases in the data: observing planets further out by the radial velocity or transit method is hard, and to observe them by direct imaging requires the planet to be bright and massive.  Given these biases and the small numbers available, it is premature to draw strong statistical conclusions about the mass ordering of planetary systems. However, the Big-Small-Big mass ordering as considered in this paper is present in a substantial fraction of the current sample, supporting research into the theoretical framework behind such systems.  It is also worth noting that our own Solar System also contains sandwiched planets (Mars and Uranus).

\subsection{Caveats}
\label{sec:caveats}

\begin{figure}
    \centering
    \includegraphics[width=\columnwidth]{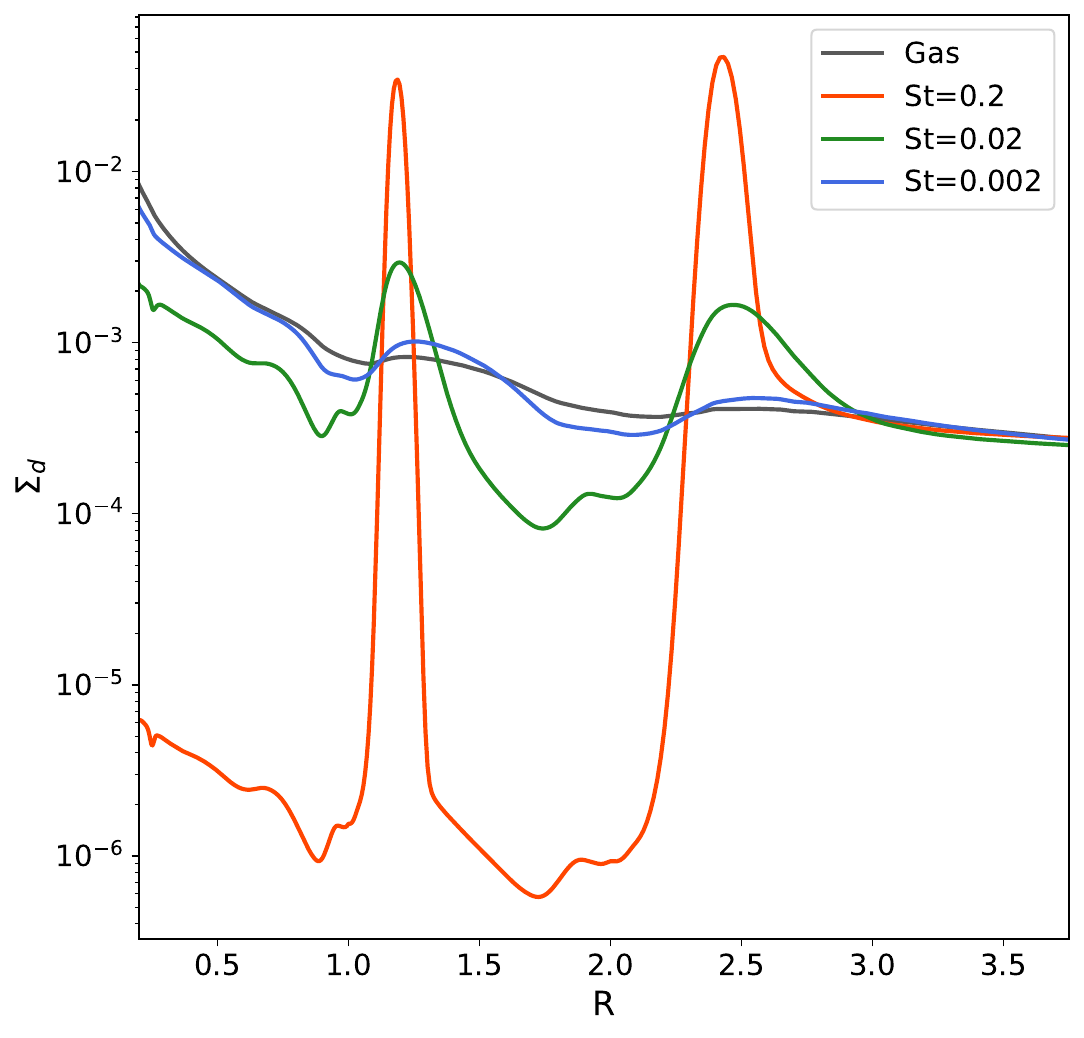}
    \caption{Azimuthally averaged dust surface density profiles for the three simulated dust species and compared to the gas surface density (where the dust is rescaled by our assumed dust-to-gas ratio of 0.01), for the $20M_{\oplus}$ and $35M_{\oplus}$ (Big-Big) system.  The small dust tends to mostly follow the gas, whilst the larger dust experiences radial drift and dust trapping, resulting in large scale dust depletion in the interplanetary region and the inner disc.}
    \label{fig:dusts}
\end{figure}

In this paper we introduce the idea of sequential planet formation happening in between two already formed planets, rather than occurring from the inner to the outer disc.  In this preliminary study we focus on the qualitative results but there are many simplifications that have been made that would need to be addressed in future studies to produce more quantitative findings.\\

\emph{Dust coagulation and fragmentation:}  This is expected to alter the dust size distribution \citep{Drazkowska_hydro_coag,Laune_hydro_coag,Lau_pressurebump_core}, with the smallest dust in general coagulating to form larger dust particles. In our two-planet simulations, while we see large scale depletions in the largest dust in the interplanetary regions, the depletion in the smallest dust is far less.  This is illustrated in Figure~\ref{fig:dusts}, where the dust density profiles of the three dust species are plotted on the same axis for the $20M_{\oplus}$ and $35M_{\oplus}$ system (and are multiplied by 100 for ease of comparison with the gas density).  We expect that the small dust could coagulate to form larger dust that is susceptible to radial drift and trapping in the inner pressure bump. It is therefore expected that accounting for coagulation could result in a larger amount of material in the interplanetary region than that presented here. Further simulations modelling dust coagulation and fragmentation would be needed to check to what extent this process would alter our conclusions.\\

\emph{Dust back-reaction}: This is expected to result in the deformation of pressure bumps when the dust-to-gas ratio exceeds unity \citep[][]{Taki2016}.  This deformation acts to smooth out pressure bumps, spreading the dust ring and reducing dust trapping effects \citep{Kanagawa_BR_dust_rings}, but does not completely erase the pressure bump \citep{Onishi_BR_dust_rings} and can still lead to planetesimal formation \citep{Carrera_PF_rings}.  In our simulations, the dust-to-gas ratio does reach the levels required for back-reaction to occur.  Therefore we would expect a different disc dust profile and likely a lesser degree of dust accumulation. 
 The general effects of radial drift and dust trapped in the outer pressure bump being unable to get to the inner bump are, however, expected to still hold. As a result, we expect the general conclusion that the amount of dust trapped in the pressure maximum in between the two planets being reduced, and thus the possibility of a low mass planet being sandwiched in between two more massive planets, to still hold.  Again, further simulations would be needed to confirm this.\\

\emph{Planetary dynamics:} In this investigation we adopt highly simplified planetary dynamics. We have not included planet migration nor planet-planet interactions, and it is certainly known that the effect of the former causes complex dust dynamics to take place \citep{Dong_migration_dust,Meru2019,Wafflard-Fernandez_Baruteau_migration_dust,Chametla_Chrenko2022}.  Furthermore, as planets migrate they can pass through strong resonances which may alter their dynamics and stability \citep[e.g.][]{Papaloizou_resonance,Quillen_Kepler36} and it is also important to consider the long-term dynamically stability of the resulting planetary system (after the disc has dispersed).  While it is clear that these effects have a significant influence on how a planetary system evolves, it is unclear how these could affect the behaviour we isolate here, if at all. The large degree of complexity these effects introduce means that a careful selection of simulations would be needed to investigate this.\\

\emph{Planet growth:} We make no assumptions about the growth of the two planets simulated.  That is, we effectively assume that they both form quickly and simultaneously. In reality a growing planet would pass from the "Small" to the "Big" regime, i.e. there would be a period where the growth of two planets in a disc may start off in the regime described by the leftmost panel of Figure~\ref{fig:schematic}, then it would transition to either of the middle panels of Figure~\ref{fig:schematic}, before potentially ending up in that described by the right panel.  Therefore there would be a period where dust is able to flow past the outer planet.  The degree to which this would collect in between the two planets would depend on the state of the inner planet at that time; for instance, if the inner planet had grown large enough to form a pressure maximum by this time, we would expect a greater degree of dust collection in the inter planetary region.  In addition we note that whilst we perform simulations of planets for 3000 orbits (which is equivalent to $\approx 9 \times 10^4$~yrs at 10au), there would continue to be a supply of material at least to the outer planet from the outer disc -- the longer term impact of this on the growth of the outer planet would need to be considered.  To determine the degree to which our conclusions would change when accounting for planetary formation and prior disc evolution, one would need to self-consistently account for when each planet would start to form as well as the formation timescales at each location.\\

\emph{Disc viscosity:} Our main set of simulations consider $\alpha=10^{-3}$. This is a commonly chosen value for hydrodynamical protoplanetary disc simulations and a reasonable estimate for typical protoplanetary discs.  Whilst in principle $\alpha$ may be as large as $\sim 10^{-2}$, there is growing evidence for lower disc viscosities ($\alpha < 10^{-4}$ \citealp{Fidele2018}; also see \citealp{Rosotti_turb_review} for a recent review). In Appendix \ref{sec:viscosity} we also consider the case of $\alpha=10^{-4}$. We draw the same general conclusions from these simulations, i.e. that a smaller mass object may form in between two more massive planets, except we find that the planet masses for which this sandwiched effect occurs are smaller.\\

\emph{Stellar and planet masses:} We note that here we have given planetary masses when assuming that the central star mass is $1M_{\odot}$.  While most stars are lower in mass, planets have been observed around both more and less massive stars.  Since {\sc fargo3d} is scale-free and works with planet-star mass ratios rather than masses themselves, one can easily rescale for other mass stars, e.g. a 20$M_{\oplus}$ planet here would correspond to a $40M_{\oplus}$ planet with a $2M_{\odot}$ host star. However, we also note that for low mass stars dust is expected to settle to the disc midplane faster \citep[][]{Mulders2012} and to move towards the star at a higher radial drift velocity \citep[][]{Pinilla2013}.

Furthermore, the masses we have chosen all correspond to relatively small planets; it is natural to ask how the results would differ in the case of giant, gap-opening planets, such as those in the PDS 70 system \citep[][]{Keppler2018, Haffert2019}. In this case the planets would have a much stronger influence on the gas profile, producing much wider and deeper gaps. We note that in such cases there would also be the added effects of gap merging/overlap to consider in addition to the effects of radial drift and dust supply cut-off.\\

\emph{Sandwiched planet, dwarf planet or moon mass object?:} It is important to point out that whilst we have presented the idea that a smaller mass planet may form in between two larger planets as a way to potentially describe a Big-Small-Big architecture of planetary systems, we note that we have not quantified the amount of dust that actually accumulates in the interplanetary region.  More accurate simulations are needed that take into account some of the above caveats to accurately determine how small or massive the sandwiched object would be.

\section{Conclusions}
\label{sec:conclusions}

Planets embedded in protoplanetary discs are known to cause rings of dust, and can have a significant impact on the amount of dust flowing into the inner disc.  One key aspect is whether such rings can be the sites of subsequent planet formation. In a two-planet system, the outer planet may influence the amount of solid material in the ring just exterior to the inner planet.  When the outer planet is small (i.e. is below the pebble isolation mass at its location), the point of inflection it produces in the disc's gas pressure profile has only a limited effect on dust inflow.  When the outer planet is large enough to generate a pressure maximum the inflow of (larger) dust is halted, as expected.  Of particular interest is the case when both planets are massive enough to generate pressure maxima in the disc.  In this scenario the amount of dust trapped exterior to the inner planet is greatly reduced compared to the single planet case.  As a result, if subsequent planet formation is to occur at this location, the architecture of the resulting planetary system is expected to show a small planet sandwiched in between two more massive planets.  This is not the architecture that is typically expected through conventional planet formation processes but may describe a subset of planetary systems that are observed to have the architecture where the middle planet is smaller than the inner and outer planets.

\section*{Acknowledgements}

We thank the referee whose insights improved this paper.  MP acknowledges support from The Royal Society Enhancement Award.  FM acknowledges support from The Royal Society Dorothy Hodgkin Fellowship.  S.R. acknowledges support from a Royal Society Enhancement Award, and funding from the Science \& Technology Facilities Council (STFC) through Consolidated Grant ST/W000857/1. DJA is supported by UKRI through the STFC (ST/R00384X/1) and EPSRC (EP/X027562/1). We also thank Richard Booth and Dimitri Veras for helpful discussions.  This work was performed using Orac and Avon, the HPC clusters at the University of Warwick.  This work made use of the following software: FARGO3D \citep[][]{FARGO3D}, Matplotlib \citep[][]{Matplotlib}, numpy \citep[][]{numpy}.

\section*{Data Availability}

 The data from the simulations used to create all plots in this article is available on reasonable request to the corresponding author.



\bibliographystyle{mnras}
\bibliography{example} 



\appendix

\section{Numerical tests}
\label{sec:Numerical tests}

Our simulations are run with an inner boundary at $R = 0.2$.  We note that there is a moderate over-density of gas,
rising sharply above the initial distribution close to the inner boundary.  Figure A1 shows the impact of changing the inner boundary to $R=0.5$ for the four fiducial test cases.  Moving the boundary outwards leads to an increase in the gas surface mass density and hence pressure gradient (at $R \lesssim 0.75$), which changes the radial drift of larger dust in this region and results in an artificial buildup of dust (not shown here).  The effect is minimised by moving the inner boundary as close to the star as is feasible. A logarithmic grid means that, for a given number of azimuthal grid cells, the number of radial grid cells quickly increases upon moving the inner boundary in. For this reason, the computational cost of simulations significantly increases for inner boundaries at smaller radii.  Our choice of an inner boundary at $R=0.2$ avoids the artificial dust buildup at a reasonable computational cost.

\begin{figure*}
    \centering
    \includegraphics[width=\textwidth]{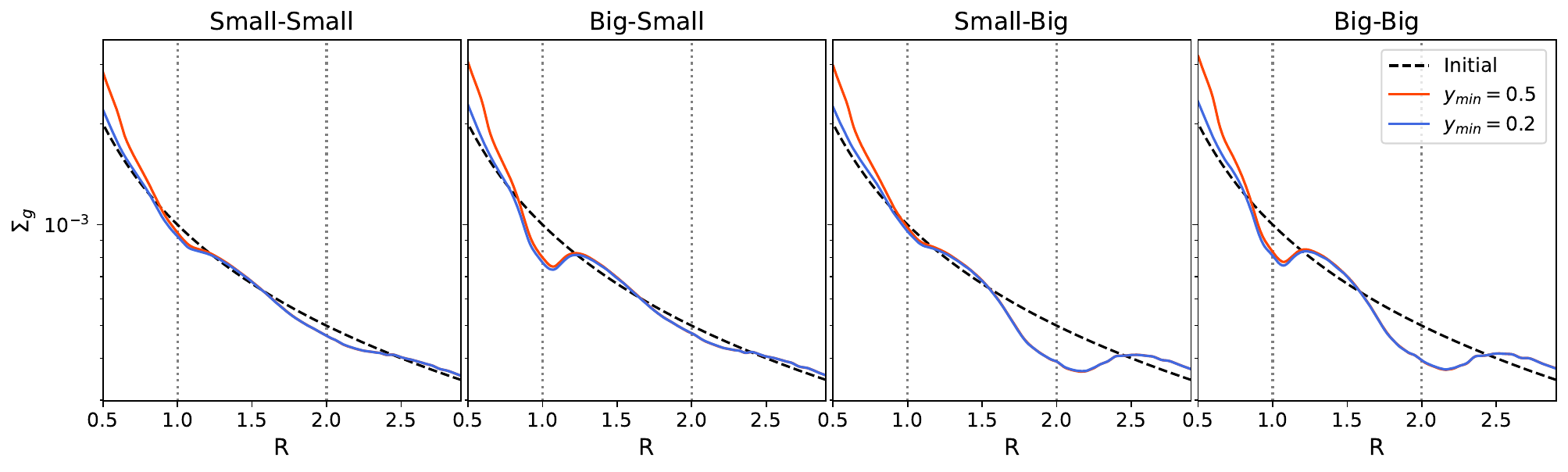}
    \caption{Azimuthally averaged gas surface density profiles for the four two-planet systems for simulations with the inner radial boundary of the computational domain located at $R = 0.5$ (red line) and $R = 0.2$ (blue line).  Moving the boundary inwards reduces the gas over-density observed within the inner disc. For $R \gtrsim 0.75$ the differences in the density profiles are minimal, resulting in our conclusions being unchanged.}
    \label{fig:ymin}
\end{figure*}

A very small gradual dip in the gas pressure profile is associated with the outer boundary. This effect only becomes noticeable in and near the outer damping zone. With the outer damping zone boundary defined as in \citet{McNally2019}, this means that this effect will only be seen in the region local to the outer planet if the outer boundary is located at $R\lesssim 4$. Hence, for the simulations presented in the main text where we set the the outer boundary to $R=10$, this outer boundary effect has no influence on our conclusions. Changing the outer boundary also affects the computational cost of a simulation, but to a lesser degree than the inner boundary.

\begin{figure}
    \centering
    \includegraphics[width=\columnwidth]{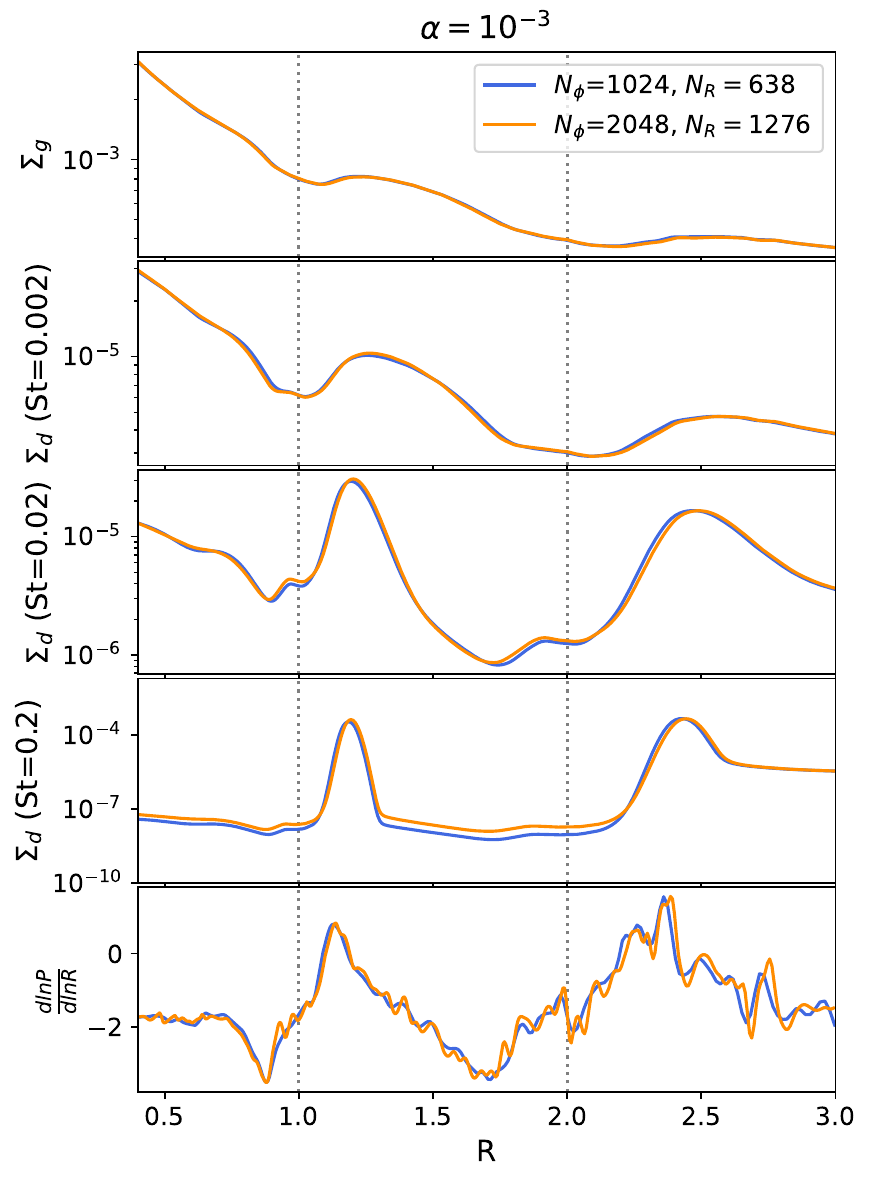}
    \caption{Disc gas and dust density and gas pressure profiles for the 20$M_{\oplus}$ \& 35$M_{\oplus}$ (Big-Big) system at two different resolutions: the resolution used in the main suite of simulations (blue line) and double resolution (orange line). Increasing the resolution has no significant impact, indicating that the results presented in Section \ref{sec:results} are robust.}
    \label{fig:restest}
\end{figure}

To test the effect of numerical resolution, we perform an additional simulation of the $20M_{\oplus}$ and $35M_{\oplus}$ system with double the resolution ($N_{\phi}=2048, N_R=1276$). The differences in the disc profiles between this case and that in the main text are minimal (see Figure \ref{fig:restest}). This increases our confidence in the validity of results of our main suite of simulations. Note that simulations at lower viscosities seem to be less robust to changes in resolution (see Appendix \ref{sec:viscosity} and Figure \ref{fig:alpharestest}).

When using similar experimental setups, some authors \citep[e.g][]{Rosotti2016} aim to reduce numerical artefacts by gradually increasing the masses of embedded planets from zero to the full planet mass over the first few tens of orbits. We did not do this for the simulations presented here. An additional test of the fiducial Big-Big case including such a mass tapering (over 20 orbits) showed that the disc profiles with and without tapering become indistinguishable beyond a few hundred orbits.

\section{Disc viscosity}
\label{sec:viscosity}

\begin{table*}
	\centering
	\caption{Table of low viscosity simulations. The simulation names correspond to the four cases of influence on the disc pressure profile illustrated in Figure~\ref{fig:schematic}. 'LV' refers to the lower viscosity ($\alpha=10^{-4}$) simulations.}
	\label{tab:LVsimulations_table}
	\begin{tabular}{lccccr} 
		\hline
		Simulation & \multicolumn{2}{c}{Inner planet}&\multicolumn{2}{c}{Outer planet}&$\alpha$\\
		& Mass ($M_{\oplus}$) & Pressure maximum? & Mass ($M_{\oplus}$) & Pressure maximum? &\\
		\hline
		\hline
		Small-Small (LV) & 6 & No & 13 & No & $10^{-4}$\\
		Big-Small (LV) & 9 & Yes & 13 & No & $10^{-4}$\\
		Small-Big (LV) & 6 & No & 18 & Yes & $10^{-4}$\\
		Big-Big (LV) & 9 & Yes & 18 & Yes & $10^{-4}$\\
		\hline
	\end{tabular}
\end{table*}

\begin{figure*}
    \centering
        \includegraphics[width=\textwidth]{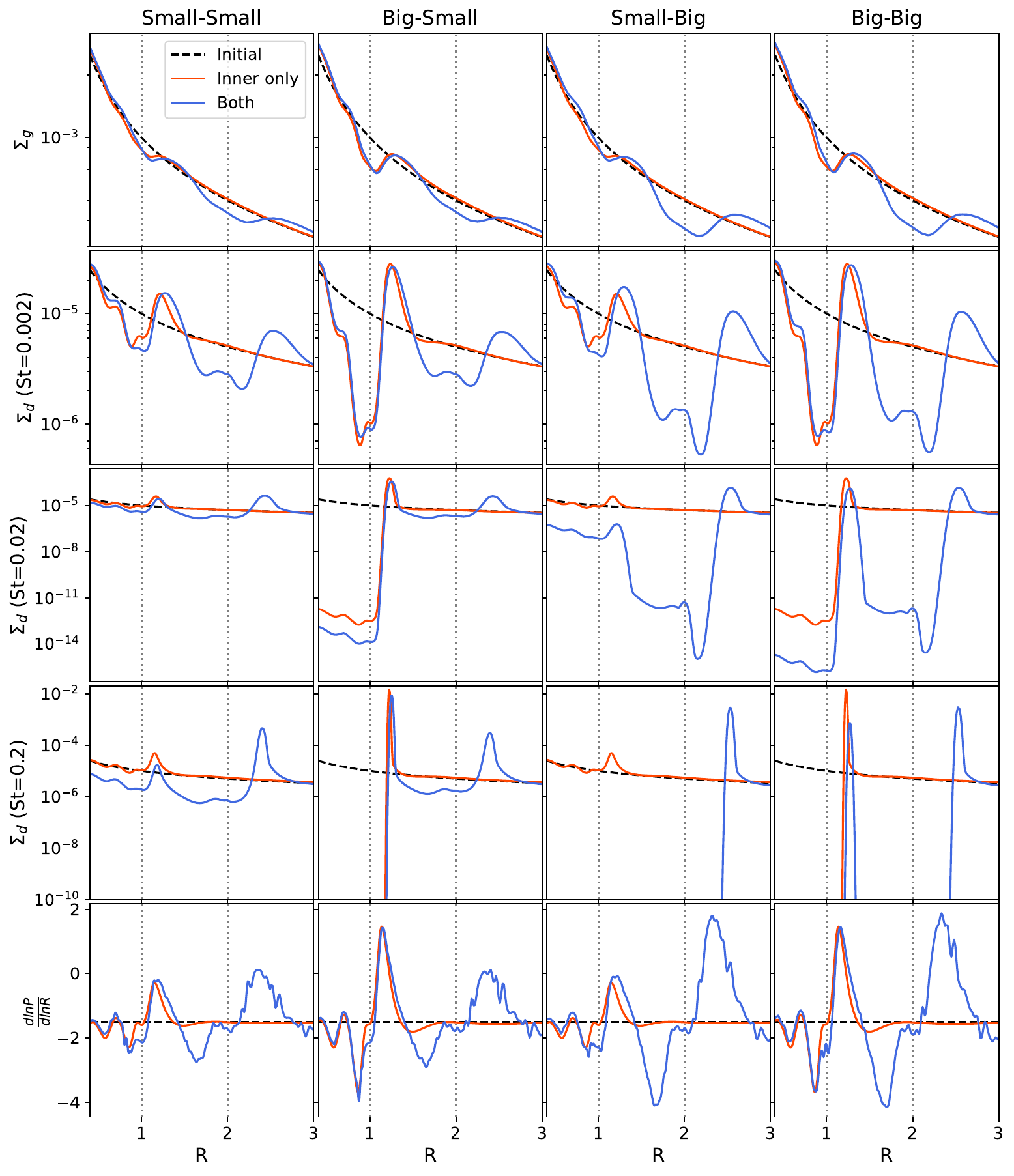}
    \caption{The equivalent of Figure ~\ref{fig:main_fig} for $\alpha=10^{-4}$. For the four cases, the planetary masses are reduced as described in Appendix \ref{sec:viscosity}. The outcomes of the four cases are similar to those for $\alpha=10^{-3}$, described in Section \ref{sec:results} in that for the case where two pressure maximum forming planets are present, the dust density in the ring exterior to the inner planet is reduced.}
    \label{fig:alphafig}
\end{figure*}

Reducing the viscosity parameter, $\alpha$, has a very large effect on the disc profile both in the the one- and two-planet cases.  Firstly, decreasing $\alpha$ reduces the isolation mass \citep[][]{Bitsch2018, Ataiee2018}.  As a result the planet masses used for the $\alpha = 1\times 10^{-3}$ disc are now able to generate pressure maxima.  Hence, less massive planets are needed to test the four cases described in Figure \ref{fig:schematic}.  We therefore choose 6 and 8$M_{\oplus}$ inner planets and 13 and 16$M_{\oplus}$ outer planets, again located at $R=1$ and $R=2$, respectively (see Table~\ref{tab:LVsimulations_table}) and run these for 9000 orbits.  The qualitative results of these simulations are similar to those described in Section \ref{sec:results}, and are displayed in Figure \ref{fig:alphafig}.  We note that the sandwiched planet formation effect discussed in Section \ref{sec:implications} still occurs for lower viscosity discs, albeit for lower planet masses.

We note that these simulations have been run for less than the viscous time.  This is purely to illustrate that the sandwiched architecture can potentially occur even for lower mass planets.  For a more quantitative study, the simulations would need to be run for longer given that the viscous time is longer in low viscosity discs.

\begin{figure}
    \centering
    \includegraphics[width=\columnwidth]{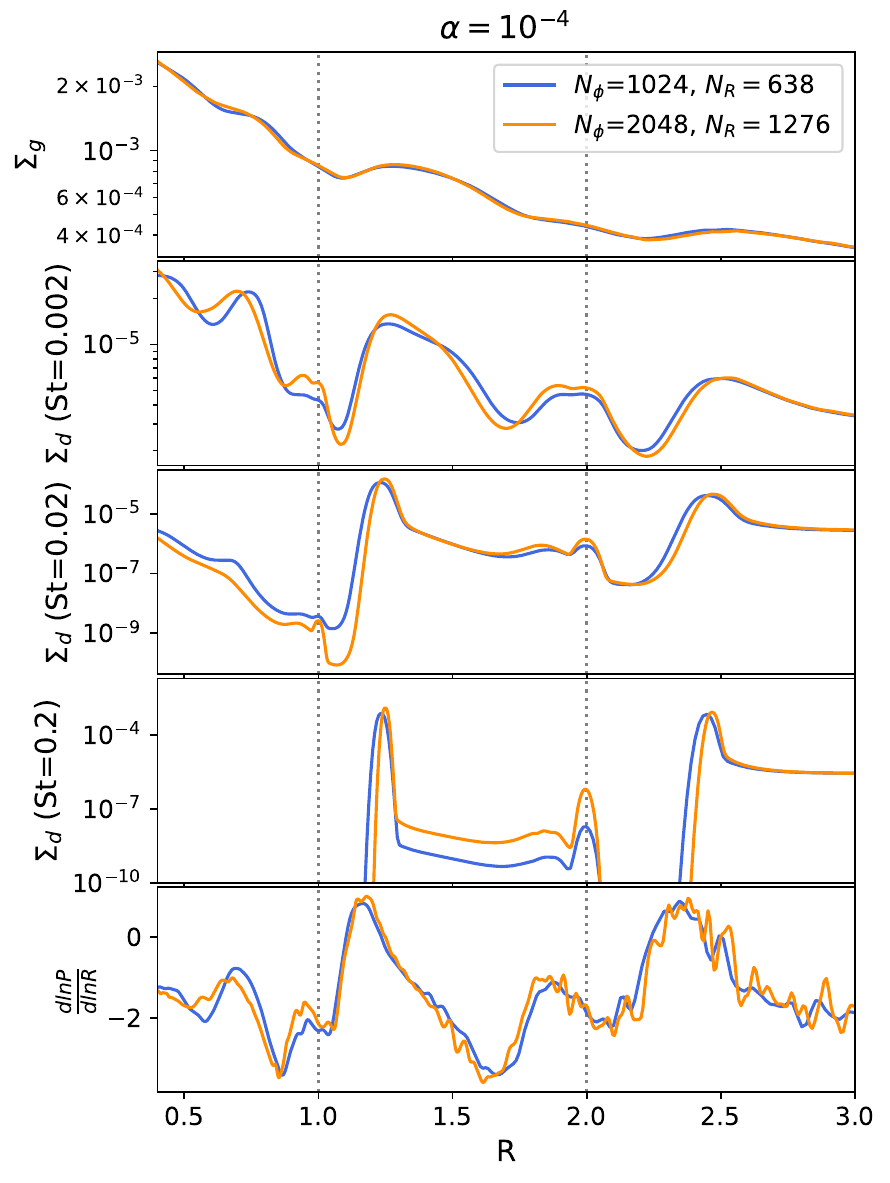}
    \caption{disc density and gas pressure profiles for the 9$M_{\oplus}$ \& 18$M_{\oplus}$ $\alpha=10^{-4}$ (Big-Big) system at $\rm t=3000$ orbits (at $\rm R=1$) at two different resolutions: the resolution used in the main suite of simulations (blue line) and double resolution (orange line). While many qualitative features are similar, notable differences are seen between the two cases. This impact shows that caution is needed when choosing the resolutions for low viscosity simulations.}
    \label{fig:alpharestest}
\end{figure}

Decreasing $\alpha$ also has an influence on the numerical artefacts discussed in Appendix \ref{sec:Numerical tests}. Firstly the artefact associated with the inner boundary is enhanced with reduced viscosity. With the inner boundary at $R=0.5$ for $\alpha = 1 \times 10^{-3}$ a large pressure peak is seen interior to the inner planet. In the case of reduced viscosity the impact of this on the disc dust profile is much larger (not shown here). Secondly, the effect of changing resolution is also more significant for these reduced $\alpha$ simulations. Whereas the effect of doubling the resolution is minimal when $\alpha=10^{-3}$ (see Figure \ref{fig:restest}), this is not the case for $\alpha=10^{-4}$; with this reduced viscosity, though most qualitative features are preserved, Figure \ref{fig:alpharestest} shows how there are significant quantitative effects when the resolution is changed. We do not closely consider the quantitative results of simulations in this paper.  Should later work wish to do so, the choice of resolution would need to be carefully considered.


\bsp	
\label{lastpage}
\end{document}